\documentclass[10pt,conference]{IEEEtran}
\usepackage{epsfig,graphicx,color}
\usepackage{times}
\newtheorem{theorem}{Theorem}

\newtheorem{proposition}{Proposition}

\newcommand{\x}{{\bf x}}
\newcommand{\E}{{\mathbf E}}
\newcommand{\rank}{{\mathrm {rank}}}
\newcommand{\GF}{{\mathrm {GF}}}
\newcommand{\C}{{\mathbf {C}}}

\begin{document}

\title{Coding Schemes for Line Networks}
\author{\authorblockN{Payam Pakzad, Christina Fragouli and Amin Shokrollahi}
\authorblockA{Laboratoire d'algorithmique (ALGO)\\
Ecole Polytechnique F\'ed\'erale de Lausanne (EPFL)\\
CH-1015 Lausanne, Switzerland \\
	\{payam.pakzad, christina.fragouli, amin.shokrollahi\}@epfl.ch}
}
\maketitle

\begin{abstract}
We consider a simple network, where a source and destination node are connected with a line of erasure channels.  It is well known that in order to achieve the min-cut capacity, the intermediate nodes are required to process the information.  We propose coding schemes for this setting, and discuss each scheme in terms of complexity, delay, achievable rate, memory requirement, and adaptability to unknown channel parameters. We also briefly discuss how these schemes can be extended to more general networks.
\end{abstract}

\section{Introduction}

Networked systems arise in various contexts such as the public internet
peer-to-peer networks, ad-hoc wireless networks, and sensor networks.
Such systems are becoming central to our everyday life. 
The networked systems today employ traditional coding schemes for end-to-end connections
and are generally not tailored to the network environment. For example,
for reasons of design simplicity, intermediate nodes at a network 
are only allowed to forward and not to process incoming information flows.
However, as the size of communication networks grows,
it becomes less clear if the benefits of the simple end-to-end approach outweigh those
of coding schemes that employ intermediate node processing.

From a theoretical point of view it is well-known that if intermediate nodes
are allowed to decode and re-encode the information sent by the source,
--with no constraints on complexity and/or delay,--
then the information capacity between a sender and a receiver is upper bounded by
the {\em min-cut capacity} of the network, as described in~\cite{Cover}.
A crucial point in making schemes that employ intermediate node-processing practical and attractive,
is in realizing benefits  without incurring excessive complexity and delay.

In this paper we propose  coding schemes that employ intermediate
node processing and discuss their performance. These  schemes 
are based on {\em fountain codes}, a set of rate-less codes recently proposed \cite{LT,Raptor}
that have a number of desirable properties for networked environments.
We compare different coding schemes based on their {\em complexity, delay, memory requirement, achievable rate,} and
{\em adaptability}; we will define these metrics precisely in Section \ref{model}.

For example, if we use an LT-code \cite{LT} 
to encode $k$ information bits at the source and simply forward any received bit at the intermediate nodes,
we would need $O(k\log(k)/C)$ XOR operations at the transmitter,
and $O(k\log(k))$  XOR operations at the receiver, where $C$ is the end-to-end capacity of the overall channel measured in bits per channel use.
Intermediate nodes would have no processing or memory requirements, and would not introduce delay.
This scheme would further adapt to unknown channel parameters.
However, the achievable rate can only approach the end-to-end capacity of the overall channel, 
which is in general less than the min-cut capacity of the network.

In \cite{isita} the authors examined the benefits of intermediate node processing from an 
information theoretic point of view. Our work can be viewed as approaching the same problem 
from a coding theory point of view.

In \cite{Lun} a scheme was proposed that takes advantage of
intermediate node processing 
to approach the min-cut capacity, and puts emphasis on the queuing theory aspects of the problem.
The authors show that if we allow intermediate nodes
to transmit random linear combinations of the incoming packets
over a finite field $\GF(q)$, the transmission rate approaches the min-cut capacity as
$q$ goes to infinity.
In this paper we will present alternative optimal coding schemes
that  approach the min-cut capacity using a constant field size,
and in particular a binary field.

The paper is organized as follows. In
Section~\ref{model} we present our model and 
performance metrics in more detail. In Section~\ref{coding} we describe our proposed coding schemes.
In Section~\ref{general} we discuss generalization to other networks; In Section~\ref{compare} we
compare our results with some related work in more details, and finally we conclude the paper in
Section~\ref{discuss}.

\section{\label{model} Model}

We consider a linear network that models a path between a source and a destination. 
The corresponding graph is comprised of a source node,
a destination node and a series of $L-1$ intermediate nodes.
The $L$ edges between the nodes correspond to independent memoryless erasure channels,
and the information units sent over the $i$th link are erased with probability $\epsilon_i$.

We assume a discrete time model, where each node can transmit one unit of information at each time slot.
For coding purposes, we will treat each information unit as a symbol, but in general we can have a packet
of symbols, and apply to each symbol of the packet the same encoding/decoding operation; in the following,
we will refer to information units as packets or symbols interchangeably. 
Intermediate nodes have the capability to process the packets they receive,
and use them to generate new packets.
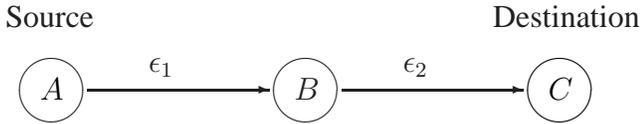
\begin{figure}[thb!]\centering
\setlength{\unitlength}{3000sp}%
\begingroup\makeatletter\ifx\SetFigFont\undefined%
\gdef\SetFigFont#1#2#3#4#5{%
  \reset@font\fontsize{#1}{#2pt}%
  \fontfamily{#3}\fontseries{#4}\fontshape{#5}%
  \selectfont}%
\fi\endgroup%

\noindent
\begin{picture}(5494,997)(800,-2813)
{\thinlines
\put(1351,-2536){\circle{540}}
}%
\put(1261,-2617){\makebox(0,0)[lb]{\smash{{\SetFigFont{12}{18.0}{\rmdefault}{\mddefault}{\updefault}{\color[rgb]{0,0,0}$A$}%
}}}}
{\color[rgb]{0,0,0}\put(5551,-2536){\circle{540}}
}%
\put(5461,-2617){\makebox(0,0)[lb]{\smash{{\SetFigFont{12}{18.0}{\rmdefault}{\mddefault}{\updefault}{\color[rgb]{0,0,0}$C$}%
}}}}
{\put(3451,-2536){\circle{540}}
}%
{\put(1651,-2536){\vector( 1, 0){1500}}
}%
{\put(3751,-2536){\vector( 1, 0){1500}}
}%
\put(3361,-2617){\makebox(0,0)[lb]{\smash{{\SetFigFont{12}{18.0}{\rmdefault}{\mddefault}{\updefault}{$B$}%
}}}}
\put(976,-2011){\makebox(0,0)[lb]{\smash{{\SetFigFont{12}{18.0}{\rmdefault}{\mddefault}{\updefault}{Source}%
}}}}
\put(5001,-2011){\makebox(0,0)[lb]{\smash{{\SetFigFont{12}{18.0}{\rmdefault}{\mddefault}{\updefault}{Destination}%
}}}}
\put(2156,-2386){\makebox(0,0)[lb]{\smash{{\SetFigFont{12}{18.0}{\rmdefault}{\mddefault}{\updefault}{$\epsilon_1$}%
}}}}
\put(4256,-2386){\makebox(0,0)[lb]{\smash{{\SetFigFont{12}{18.0}{\rmdefault}{\mddefault}{\updefault}{$\epsilon_2$}%
}}}}
\end{picture}%
\caption{\label{fig:line} A path between a source $A$ and a receiver $C$ with L=2 links.}
\end{figure}
We ignore the transmission delay along channels (as it is beyond our control),
i.e., we assume that a packet transmitted at time $d$, if not erased, is received immediately at the next node in the chain.

Throughout this paper we will use as illustrating example the simple configuration depicted in 
Fig.~\ref{fig:line} with $L=2$ links; we will also discuss the generalization of our results 
to longer chains.  
The source node $A$ encodes $k$ symbols to create $n_1$ coded outputs using a code $\C_1$ and
sends them over the channel $AB$. Node $B$ will receive on average $n_1(1-\epsilon_1)$
coded symbols over $n_1$ time slots. Node $B$ will send  $n_2$ packets, using a code (more 
generally, processing) $\C_2$.  If node $B$ finishes transmitting at time $d$, where
$\max \{n_1,n_2\} \leq d \leq n_1+n_2$, then node $C$ will receive on average 
$n_2 (1-\epsilon_2)$ packets after $d$ time slots.
For each coding scheme of this type, we define the following metrics:
\begin{enumerate}
\item {\it Complexity} for encoding/processing/decoding at nodes $A$, $B$ and $C$: the number of operations
required as a function of $k$, $n_1$ and $n_2$.
\item {\it Delay} incurred at the intermediate node $B$: this is the time $(d-k/C_{\mathrm {mc}})$, where $C_{\mathrm {mc}}$
is the min-cut capacity.  We will remark more on this notion of delay in Section~\ref{s:delay} below.
\item {\it Memory requirement}:  the number of memory elements needed at node $B$.  Section~\ref{s:delay} will also comment
on the minimal memory requirements of any coding scheme over the line network.
\item {\it Achievable rate}: the rate at which information is transmitted from $A$ to $C$.
We say that a coding scheme is optimal in rate, if  each individual link is used at a rate equal to its capacity.
Thus it can achieve the min-cut capacity
between the source and the destination.

\item {\it Adaptability}: whether the coding scheme needs to be designed for specific erasure probabilities
$\epsilon_1$ and $\epsilon_2$ or not.
Fountain codes, for example, are adaptable in this sense.
\end{enumerate}

We observe that, although it is possible to design a code over a single link that is both adaptable and 
is optimized for achievable rate and delay, the overall coding scheme cannot be adaptable if we want 
to jointly  optimize for achievable rate and delay.
Indeed, assume that $\epsilon_2=0$. Then the scheme that jointly optimizes the delay and the achievable rate
requires node $B$ to transmit (forward) only when it receives a new packet.
However, if  $\epsilon_1$ and $\epsilon_2$ are equal and large,
then a large fraction of the packets will get erased.
In order to optimize for delay,
node $B$ should transmit about $\frac{1}{1-\epsilon_2}$ packets 
for each packet it receives,
without waiting to receive the next packet from node $A$.  Therefore a single scheme cannot be rate-optimal
for both cases.

Depending on the application, different emphases might be placed on these performance metrics.
For example, consider a real-time application, where information 
is collected into blocks of $k$ packets that are encoded and sent over the channel.
In other words, we want to transmit the real-time information from a source, as it is produced.
Assume that we have $M$ such blocks. 
Then the delay overhead at intermediate nodes can be considered to be a ``set-up''
delay for the connection, experienced only once, and hence insignificant if $M$ is large.
On the other hand, the memory requirements at intermediate nodes may be restrictive.
Indeed, there might exist a large number of connections (paths) 
that share an intermediate node that performs processing.
Thus, the memory available for each individual connection might need to be scaled down accordingly.

\subsection{Optimal Delay and Memory Requirements}\label{s:delay}
Recall that our notion of delay is linked with the optimal time of communication over a {\em single} channel with equivalent min-cut
capacity.
Note however that with this definition, it is impossible to achieve a `zero delay' scheme even for the simple network of 
Fig.~\ref{fig:line}.  In fact, even if both links $AB$ and $BC$ provide perfect feedback, there is an inherent delay
to be suffered due to the existence of sequential links.  As we will see, even in this perfect setting, there is also a need 
for memory storage, in amounts that grow with $k$.
In this section we will calculate the memory requirements, as well as the minimal delay which is incurred 
when perfect feedback exists; certainly no coding scheme that does not rely on feedback can transmit in less time.

The obvious optimal scheme in the presence of feedback is one where each node repeats transmission of each packet until it is
successfully received at the destination.  Node $A$ then completes transmitting in time $n\approx k/(1-\epsilon)$.
The operations at node $B$ can be described using a Markov chain with states 
$x_i\in\{0,1,2,\cdots\}$, indicating the number of received packets still to be sent at each time; therefore at each time $i$, $x_i$ 
packets need
to be stored in memory.  At each time (when $x_i\neq 0$), with a probability $1-2\epsilon(1-\epsilon)$ the state is unchanged, and 
with a probability $2\epsilon(1-\epsilon)$, the state is increased or decreased by $1$, with equal probability.  Therefore, after
$n$ time slots, the dynamics of this system resembles that of a random walk with a reflecting boundary at $0$, over 
$n'=2\epsilon(1-\epsilon)n$ steps; (there is slight correction, due to `longer stays' at state $0$, but for large $n$, the probability
of being at that state is insignificant.)  Thus the expected value of $x_n$ is the expected value of the absolute value of a random
walk after $n'$ steps.  Therefore $\E[x_n]=O(\sqrt{n'})=O(\sqrt{2\epsilon k})$, where we have used that $n\approx k/(1-\epsilon)$.
Node $B$ then completes transmitting the remaining $x_n$ packets in a time $d\approx x_n/(1-\epsilon)$.  Therefore, the `delay' of this 
scheme is $O(\sqrt{\epsilon k}/(1-\epsilon))$, while the expected memory requirement is $O(\sqrt{\epsilon k})$.

This argument can be extended to show that in linear network with $L$ similar links, where $L$ is a fixed finite number, each 
intermediate node incurs a delay of $O(\sqrt{\epsilon k}/(1-\epsilon))$ and requires $O(\sqrt{\epsilon k})$ units of memory.

\section{\label{coding} Coding Schemes}

In this section we describe and compare a number of coding schemes
for a line network with $L$ links. In the next section we will discuss how these schemes can be extended to more general settings.

We will use the configuration in Fig.~\ref{fig:line}, with $L=2$, as the illustrating example, and assume
for simplicity that $\epsilon_1=\epsilon_2=:\epsilon$, in which case $n_1=n_2=:n$.
In all schemes below we will use as code $\C_1$ over the link $AB$,
a fountain code, such as an LT-code or a Raptor code; as demonstrated in \cite{LT} and \cite{Raptor}, these 
codes are low complexity, rate optimal, adaptable codes over erasure channels.
Then for each different coding scheme, we will specify the code $\C_2$ over the link $BC$.
A summary of the properties of all these schemes will be provided in Table~\ref{tabl1}.

\subsection{Complete Decoding and Re-encoding}
An obvious scheme is to use a separate  code for each of the $L$ links of the line
network, and have each intermediate node completely decode and re-encode the incoming data.
Then it is obvious that we can achieve the min-cut capacity by using optimal codes (e.g. LT-codes) over each link.
However, the system suffers a delay of about $k \epsilon/(1-\epsilon)$ time-slots due to each intermediate node.
Indeed, at node $B$, we can directly forward the $(1-\epsilon)n$ received coded bits without delay,
and then, after decoding, create and send an additional  $\epsilon n$ bits over the second channel.

This straightforward scheme imposes low complexity requirements.
We only need $O(k\log(k))$ binary operations at each intermediate node to decode and re-encode an LT-code,
and the complete decoding and re-encoding scheme has memory requirements of the order $O(k)$. 
Moreover, LT-codes adapt to unknown channels in the sense defined previously.

\subsection{Systematic Codes}
The complete decoding and re-encoding scheme of the previous section is adaptable, rate optimal and has low complexity.
However it requires each intermediate node to store in memory the entire $k$ packets of information in order to re-encode.
We propose a class of coding schemes, which we call {\em systematic schemes}, which minimize the memory requirement
at the intermediate nodes, but require the knowledge of the erasure probabilities of the links.

Once again we consider the network in Fig.~\ref{fig:line} and assume that we use a fountain code $\C_1$
for link $AB$.  
In a systematic scheme, the intermediate node $B$ first forwards each coded bit (packet) from $\C_1$ as they are received; 
these are the systematic bits (packets).
Meanwhile, $B$ forms (about) $n\epsilon=\frac{k \epsilon}{1-\epsilon}$ linear combinations of the systematic bits, which 
are transmitted in the $n\epsilon$ time slots following the transmission of the systematic bits.  
Thus all systematic codes will incur an average delay of $\epsilon n$,
and will require $\epsilon n$ memory elements.  The savings in memory, as compared to the complete decoding and re-encoding,
is significant when the erasure probability $\epsilon$ is small.

In a linear network with $L$ links, the same scheme can be repeated
at each intermediate node.  Since the operation at each intermediate node is rate-optimal, it follows that for each
fixed $L$, the overall end-to-end transmission is also rate-optimal for large enough block length $k$, while each
intermediate node requires about $\epsilon n$ memory elements and contributes a delay of $n/(1-\epsilon)$.

Below we will discuss a few possible methods to design systematic codes.

\subsubsection{Fixed Codes}
Here we use a fixed systematic code, consisting of $k$ systematic
bits (packets) and $k\epsilon/(1-\epsilon)$ parity coded bits, to transmit the information over link $BC$.
A systematic LT-code \cite{Raptor}, or a Tornado code \cite{Tornado}, for example, can be used to generate the parity bits, 
and in fact any fixed systematic code can be used for this purpose.
Although not adaptable to unknown channel parameters, these codes have very low encoding and decoding complexities.  
Tornado codes for example can be encoded and decoded with $O(n\log(1/\delta))$ operations, 
where $\delta$ is a constant expressing the (fixed) rate penalty.

\subsubsection{Sparse Random Codes}
In this scheme, the non-systematic packets are formed as random (sparse) linear combinations of the
systematic ones.  More precisely, whenever a new packet is received at $B$, it is added to the storage 
space allocated to each of the non-systematic packets independently and with a (small) probability $p$.

\begin{theorem}\label{t:sysiid}
With $p=(1+\delta)\log(\epsilon k)/(\epsilon k)$ for \mbox{$\delta>0$}, the described systematic random code 
asymptotically achieves the capacity over the channel $BC$.  
\end{theorem}

\begin{proof}[Sketch]
Suppose $k'\approx k(1-\epsilon)$ systematic symbols are received at $C$, and let $l=k-k'\approx \epsilon k$.  We will then
wait for a further $l+c\log_2(l)$ non-systematic symbols to be also received at $C$, where $c>1$ is a constant.
After eliminating the received systematic symbols, these linear combinations can be described by a 
random $(l+c\log(l))\times l$ binary matrix, with i.i.d. entries which are nonzero with probability 
$p=(1+\delta)\log(\epsilon k)/(\epsilon k)$.  The results of \cite{Bloemer} can be extended to show that,
if \mbox{$p>\log(l)/l$}, the probability that such a matrix is not full-rank
approaches zero polynomially fast with $l$.
Using this and the law of large of numbers then, with high probability $C$ can retrieve all the $k$
symbols received at $B$, --e.g. by applying Gaussian elimination to this sparse matrix,-- which can then be used
to decode the fountain code $\C_1$.  This code can decode the $k$ information symbols from an average of $k+c\log(k\epsilon)$ received
symbols at $C$, and hence this scheme rate optimal for large $k$.
\end{proof}

The complexity of decoding this code is that of inverting the sparse $k\epsilon\times k\epsilon$ matrix, which is
$O((k\epsilon)^2\log(k\epsilon))$.
In fact, it can be shown that $O(\log(k)/k)$ is the smallest possible value for the probability $p$, and equivalently the density 
of the non-systematic part of the code, if the code is to be decodable with negligible overhead.  In that sense, the scheme 
provided here offers the lowest decoding complexity for any such random code where the parity bits are chosen as linear combinations 
of the systematic bits with i.i.d. distribution.

\begin{table*}[t]
\renewcommand{\arraystretch}{1.2}
\caption {\label{tabl1} Coding schemes that send $k$ bits from the source to the destination.}
\begin{center}
\begin{tabular}{|c|c|c|c|c|c|}\hline
Scheme & Intermed. node complexity & Delay & Memory & Adaptable & Rate Optimal\\\hline
Optimal (Feedback) &  $0$ & $\sqrt{k\epsilon}/(1-\epsilon)$  & $\sqrt{k\epsilon}$  & yes & yes \\\hline
Complete Dec-Reenc &  $k\log k/(1-\epsilon)$ & $k\epsilon/(1-\epsilon)$  & $k$  & yes & yes \\\hline
Systematic Fixed & $k\log(1/\delta)/(1-\epsilon)$ & $k\epsilon/(1-\epsilon)$  & $k\epsilon$  & no & yes \\\hline
Systematic Random  & $(k\epsilon)^2\log(k\epsilon)$ & $k\epsilon/(1-\epsilon)$  & $k\epsilon$  & no & yes \\\hline
Greedy Random & $k^2\log(k)$ & $\sqrt{k\epsilon\log(k\epsilon)}/(1-\epsilon)$  & $k$  & yes & yes \\\hline
\end{tabular}
\end{center}
\end{table*}

\subsection{Greedy Random Codes}\label{s:greedy}
In this scheme, at each time slot the intermediate node $B$ transmits random linear combinations (over $\GF(2)$) of 
all the packets it has received thus far.  

The main advantages of this random scheme are its adaptability and optimality in terms of delay.  The drawbacks are 
large memory requirement, and high decoding complexity, which is $O(k^2 \log k)$ XOR operations on packets.

We will need the following proposition to analyze the optimality of greedy random codes.

\begin{proposition}\label{p:rtriang}
Given a constant $c>1$, let $A$ be a `random lower-triangular' $(k+c\log(k))\times k$ binary matrix, where 
the entries $A_{i,j}$ are zero for $1\leq i<j\leq k$, and all the other entries are 
i.i.d. ${\mathrm {Bernoulli}}(1/2)$ random variables.  Then  
\[\Pr\big[\rank(A)< k\big]\leq \frac{1}{2 k^{c-1}}.\]
\end{proposition}
\begin{proof}
Let $K$ denote the right kernel of $A$, i.e., 
\[K:=\{\x\in\GF(2)^k\;|\;A\cdot \x=0\}.\]
We will find the expected size of $K$.  Let 
\[V_i:=\{\x\in\GF(2)^k\,|\,x_i=1,\;\textup{and for $j<i$}\;x_j=0\},\] 
that is,  $V_i$ is the set of 
vectors which have their first $1$-components at position $i$; then are $2^{k-i}$ such vectors.  
Let $A_j$ denote the $j$th row of $A$.  Then it is easy to verify that, for any $\x\in V_i$, 
the probability that $A_j\cdot\x=0$ is one for 
$j<i$, and is $1/2$ for $j\geq i$.  Therefore the expected size of the intersection of $V_i$ and
$K$ is \[2^{k-i}\cdot (\frac{1}{2})^{k+c\log(k)-i+1}=\frac{1}{2 k^c}.\]  The sets $V_i$ for $i=1,\cdots,k$
partition $\GF(2)^k\backslash\{0\}$, thus the expected size of $K$ is
\begin{equation}\label{e:ek}\E[|K|]=1+\sum_{i=1}^k \frac{1}{2 k^c}=1+\frac{1}{2 k^{c-1}}.\end{equation}
Now the expected size of the kernel can be used to bound the probability that $A$ is not full-rank:
\begin{eqnarray*}
\E[|K|]&=\sum_{i=0}^k \Pr[\rank(A)=k-i] 2^i\\
	& \geq \Pr[\rank(A)=k]+2\Pr[\rank(A)<k]
\end{eqnarray*}
It follows that $\Pr[\rank(A)<k]\leq \E[|K|]-1 = \frac{1}{2 k^{c-1}}.$

\end{proof}

An immediate consequence of Proposition~\ref{p:rtriang} is that, if the channels were noiseless, i.e., $\epsilon=0$,
then the greedy random coding scheme described above is rate optimal; this is because, with high probability, node $C$ 
can perform Gaussian elimination on the generator matrix of the code $\C_2$, which is a random lower-triangular matrix 
of the type discussed in Proposition~\ref{p:rtriang}.

A closer examination of the proof of Proposition~\ref{p:rtriang} reveals that, in order to make $\E[|K|]-1$ converge to zero,
it is sufficient that, for each column $i$, the matrix $A$ contains at least $k+c\log(k)-i$ rows with ${\mathrm{Bernoulli}}(1/2)$
random variables at the $i$th position;  this will then guarantee that the size of $K\cap V_i$ is no more than $1/k^c$, for some 
$c>1$, and we use (\ref{e:ek}) to obtain the desired result.
The interpretation of this statement in the context of our coding scheme is that, in order to be able to decode with high probability
at $C$, it is sufficient that for each $i=1,\cdots, k$, at least  $k+c\log(k)-i$ packets are successfully transmitted over $BC$ after
$B$ has received the $i$th coded packet from $A$.  

Let $\alpha_d$ and $\beta_d$ denote the number of packets successfully transmitted over links $AB$ and $BC$ respectively.
Suppose now that we end transmission at a time $n$ when $C$ has received $k+l$ packets, i.e., $\beta_n=k+l$, where $l=o(k)$ will be
appropriately chosen.  Then the number of packets that will be received by $C$ after a time $d$ is equal to $k+l-\beta_d$; this,  
we would like to be at least $k+c\log(k)-\alpha_d$.  In other words, the sufficient conditions above require that 
at each time $d=1,\cdots,n$, the quantity $\alpha_d-\beta_d$ be greater than $c\log(k)-l$.  But $x_d:=\alpha_d-\beta_d$ behaves 
similar to a symmetric one-dimensional random walk:  in fact, in $1-2\epsilon(1-\epsilon)$ fraction of the time slots, 
$x_d$ remains unchanged, while in the other $2\epsilon(1-\epsilon)$ fraction, it increases or decreases by $1$ with probability $1/2$.
Therefore, in $n\approx (k+l)/(1-\epsilon)$ time it takes to complete transmission as described above, $x_d$'s movements are identical
to $n'$ steps of a random walk $\{y_i\}$, where $n'=2n \epsilon(1-\epsilon)\approx 2 k\epsilon$.  Straightforward calculation then shows 
that with $l=O(\sqrt{n'\log(n')})=O(\sqrt{k\epsilon\log(k\epsilon)})$, the probability that $\{y_i\}$ at any time $i\le n'$ goes below 
$-l$ is polynomially small in $k$.  This proves that, with high probability,  the $k$ packets of information can be 
retrieved at $C$ from 
$k(1+\sqrt{\frac{\epsilon \log(k\epsilon)}{k}})$ received packets.  The overhead goes to zero as $k$ becomes large, 
and hence this coding scheme is asymptotically rate optimal.

\section{\label{general} General Networks}
In this section, we will 
represent a  communication network of binary erasure channels 
as a directed acyclic graph.

Assume for simplicity that all edges of the graph have the same capacity $C_0$.
Consider a unicast connection; then the min-cut capacity between the source and the destination is $m C_0$ for some integer $m$. 
It is straightforward to see that if we are employing a capacity-achieving coding scheme, 
it is sufficient to route the information along $m$ parallel paths $P_1,\cdots, P_m$,
where each path $P_i$ consists of $L_i$ links. We can then directly apply the coding
schemes previously described to each path separately.

In practice, since coding schemes will employ codewords of
finite block lengths, there might exist benefits 
in combining independent information streams \cite{isit}.
Moreover, not all edges might be used at the same rate,
for example because of cost considerations.

Consider a routing scheme that observes the flow conservation principle
and utilizes each edge at rate smaller or equal to its capacity.  Since all the component codes are linear, 
the received symbols along a link $l$ in the network can be described using an $(n_l \times k_l)$ matrix, 
where $k_l$ is the number of information symbols sent along the link, and $n_l$ is the number of received symbols.
The point we make in this section is that,
as long as all such matrices corresponding to the intermediate links have full column rank,
the end-to-end  matrix that the receiver will have 
to decode in order to retrieve the information bits, will also be full rank and hence decodable.

\begin{figure}[thb!]
\centering
\setlength{\unitlength}{3500sp}%
\begingroup\makeatletter\ifx\SetFigFont\undefined%
\gdef\SetFigFont#1#2#3#4#5{%
  \reset@font\fontsize{#1}{#2pt}%
  \fontfamily{#3}\fontseries{#4}\fontshape{#5}%
  \selectfont}%
\fi\endgroup%
\begin{picture}(4299,2411)(4114,-6368)
{\color[rgb]{0,0,0}\thinlines
\put(5242,-5152){\circle{450}}
}%
{\color[rgb]{0,0,0}\put(6284,-6135){\circle{450}}
}%
{\color[rgb]{0,0,0}\put(6289,-4190){\circle{450}}
}%
{\color[rgb]{0,0,0}\put(7289,-5161){\circle{450}}
}%
{\color[rgb]{0,0,0}\put(4126,-5161){\vector( 1, 0){900}}
}%
{\color[rgb]{0,0,0}\put(5401,-5011){\vector( 1, 1){675}}
}%
{\color[rgb]{0,0,0}\put(5401,-5311){\vector( 1,-1){675}}
}%
{\color[rgb]{0,0,0}\put(6451,-5986){\vector( 1, 1){675}}
}%
{\color[rgb]{0,0,0}\put(6451,-4336){\vector( 1,-1){675}}
}%
{\color[rgb]{0,0,0}\put(7501,-5161){\vector( 1, 0){900}}
}%
\put(5156,-5210){\makebox(0,0)[lb]{\smash{{\SetFigFont{10}{14.4}{\rmdefault}{\mddefault}{\updefault}{\color[rgb]{0,0,0}$A$}%
}}}}
\put(6206,-4261){\makebox(0,0)[lb]{\smash{{\SetFigFont{10}{14.4}{\rmdefault}{\mddefault}{\updefault}{\color[rgb]{0,0,0}$B$}%
}}}}
\put(6206,-6211){\makebox(0,0)[lb]{\smash{{\SetFigFont{10}{14.4}{\rmdefault}{\mddefault}{\updefault}{\color[rgb]{0,0,0}$C$}%
}}}}
\put(7201,-5210){\makebox(0,0)[lb]{\smash{{\SetFigFont{10}{14.4}{\rmdefault}{\mddefault}{\updefault}{\color[rgb]{0,0,0}$D$}%
}}}}
\put(5501,-4636){\makebox(0,0)[lb]{\smash{{\SetFigFont{10}{14.4}{\rmdefault}{\mddefault}{\updefault}{\color[rgb]{0,0,0}$R_1$}%
}}}}
\put(5501,-5800){\makebox(0,0)[lb]{\smash{{\SetFigFont{10}{14.4}{\rmdefault}{\mddefault}{\updefault}{\color[rgb]{0,0,0}$R_2$}%
}}}}
\put(6826,-4636){\makebox(0,0)[lb]{\smash{{\SetFigFont{10}{14.4}{\rmdefault}{\mddefault}{\updefault}{\color[rgb]{0,0,0}$R_1$}%
}}}}
\put(6826,-5800){\makebox(0,0)[lb]{\smash{{\SetFigFont{10}{14.4}{\rmdefault}{\mddefault}{\updefault}{\color[rgb]{0,0,0}$R_2$}%
}}}}
\put(7551,-5086){\makebox(0,0)[lb]{\smash{{\SetFigFont{10}{14.4}{\rmdefault}{\mddefault}{\updefault}{\color[rgb]{0,0,0}$R_1+R_2$}%
}}}}
\put(4201,-5086){\makebox(0,0)[lb]{\smash{{\SetFigFont{10}{14.4}{\rmdefault}{\mddefault}{\updefault}{\color[rgb]{0,0,0}$R_1+R_2$}%
}}}}
\end{picture}%
\caption{\label{fig:parallel} Splitting and merging of information in the network.}
\end{figure}
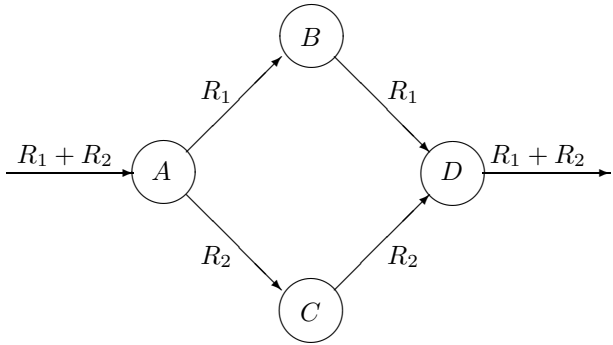

Indeed, given the matrices associated with all individual
links, to create the end-to-end  matrix, we will have to perform the following types of matrix operations:
\begin{itemize}
\item Partitioning a matrix into parts, to create the equivalent matrix corresponding to splitting an input stream to multiple
 outgoing streams, such as node $A$ in Fig.~\ref{fig:parallel}.
\item Multiplication of matrices, in order to create the equivalent matrix corresponding 
to serially concatenated channels, such as nodes $B$ and $C$ in Fig.~\ref{fig:parallel}.
\item Finding the direct sum of matrices, to create the equivalent matrix corresponding to merging multiple input streams of a 
node into a single outgoing stream, such as node $D$ in Fig.~\ref{fig:parallel}.
\end{itemize}
All these operations preserve the full-rank property.
Thus, all coding schemes described in Section~\ref{coding} can be directly applied
over a more general network.
However, for this general case, a thorough study of the delay and memory requirements for each scheme is not provided here.

\section{Comparison with previous work}\label{compare}

In \cite{Lun}  a scheme was proposed that takes advantage of
intermediate node processing 
to approach the min-cut capacity. 
The authors model the departures and arrivals at nodes as Poisson  processes
and work out the queuing-theory aspects of the problem.
The employed coding scheme allows intermediate nodes
to transmit random linear combinations of the incoming packets
over a finite field $\GF(q)$. 
The transmission rate approaches the min-cut capacity as $q$ goes to infinity.
This scheme, as described in \cite{Lun}, requires
 $O(k^2(1-\frac{1}{q}))$ operations to encode $k$ symbols at the transmitter,
$O(k^3)$ operations for decoding at the receiver,
and  $O(k^2(1-\frac{1}{q}))$ operations at each intermediate node.
Moreover, the operations are over $\GF(q)$ that are more complex than binary operations.
Intermediate nodes require storage capabilities for $k$ packets over $\GF(q)$.

The main benefit of the scheme in \cite{Lun} is in terms of delay
as we do not decode at each intermediate node.
Indeed, complete decoding and re-encoding
requires a delay of $\epsilon n$ time-slots.
However, note that the scheme in \cite{Lun}
achieves the min-cut rate for large $q$, i.e., assuming that we
are able to send $\log_2(q)$ bits per time-slot instead of one bit per time-slot
as we assume. Thus in this sense it is not clear that the comparison is fair.

In fact, the coding scheme employed in \cite{Lun}  can be thought
as employing the greedy random codes in Section~\ref{s:greedy},
where the linear combinations are performed over $\GF(q)$ instead of the binary field,
and where the encoding matrix is not sparse.
Thus our results can be viewed as an improvement over the coding scheme proposed in \cite{Lun}.

\section{\label{discuss} Conclusion}
In this paper we have examined the problem of communication over a line network, where processing of information at the
intermediate nodes is required in order to achieve the min-cut capacity, and we have included guidelines to 
extend our results to general networks.
We have proposed coding schemes based on fountain codes.
Each scheme has been analyzed and evaluated in terms of 
complexity, delay, memory requirement, achievable rate, and adaptability (see Table~\ref{tabl1}).
In general, there is a trade-off between these desirable properties, and an absolute best scheme is not claimed.

\end{document}